\documentclass[journal]{IEEEtran}
\usepackage{amssymb}
\usepackage{textgreek}

\usepackage{adjustbox}

\usepackage{cite}
\usepackage{hyperref}

\ifCLASSINFOpdf
  \usepackage{graphicx}
  \usepackage{caption}
  
\else
\fi
\usepackage{amssymb}
\usepackage{amsmath}
\usepackage{etoolbox}
\usepackage{algorithm,algorithmicx,algpseudocode}
\usepackage[colorinlistoftodos]{todonotes}
\usepackage{multicol}
\usepackage{pgfplots}
\usepackage{tikz}
\usepackage{subcaption}
\usepackage{multirow}
\begin{document}
%
\title{Adaptive Cost Coefficient Identification for  Optimal Operation in Automatic Internal Transportation}
%
%
%

\author{Pragna~Das, 
        Llu{\'\i}s~Ribas-Xirgo,~\IEEEmembership{Member,~IEEE}
        
\thanks{P Das is with Microelectronics and Systems Electronics Department, Autonomous University of Barcelona, Bellaterra, Barcelona, 08193 Spain e-mail: (pragna.das@uab.cat).}
\thanks{L. Xirgo is with Microelectronics and Systems Electronics Department, Autonomous University of Barcelona, Bellaterra, Barcelona, 08193 Spain e-mail: (lluis.ribas@uab.cat).}}
\maketitle

\begin{abstract}
The states of batteries and environment are changing always in an automated logistics. In case of planning, the current state of mobile robots (MRs) and environment plays crucial role. Thus, decisions in MRs need to incorporate the real-time states of battery, environment, \textit{et~cetera}. These states contribute to form the cost of performance of tasks. The usual practice is to use heuristics cost for optimal planning. However, the true cost is not accounted through this. A new method to compute these cost parameters, based on state of charge of batteries and environment is proposed. A scaled prototype platform is developed and topology map is used to describe the floor. The MRs traverse different paths to carry materials. The travel time is identified as the key parameter to understand costs of performances on different states of battery and floor. With suitable predictions of these travel times the cost involved to traverse from one node to another can be known. The travel times are timed linked to each arc of the map. Suitable state-space model is formulated and Kalman filtering is used to estimate these travel time online. Dijkstra's algorithm is modified to incorporate travel time as edge costs in route planning to generate paths with minimal cost. The paths are computed constantly and average of total path costs of these paths are compared with that of paths obtained by heuristics costs. The results show that average total path costs of paths obtained through on-line estimated travel times are 15\% less that of paths obtained by heuristics costs.
\end{abstract}
\IEEEpeerreviewmaketitle
\section{Introduction}
\label{intro}
\IEEEPARstart{M}{obile robot} (MR) based systems used for internal logistics in factories demand cost efficient decisions on planning and co-ordination. Usually, the information about current condition of robot, floor, batteries and other robots play crucial role in decision making \cite{colby2015implicit, gerkey2004explicit, bayram2016coalition}. 
This idea is explained in the following example. 
\begin{figure}[h]
\centering
\includegraphics[scale = 0.36]{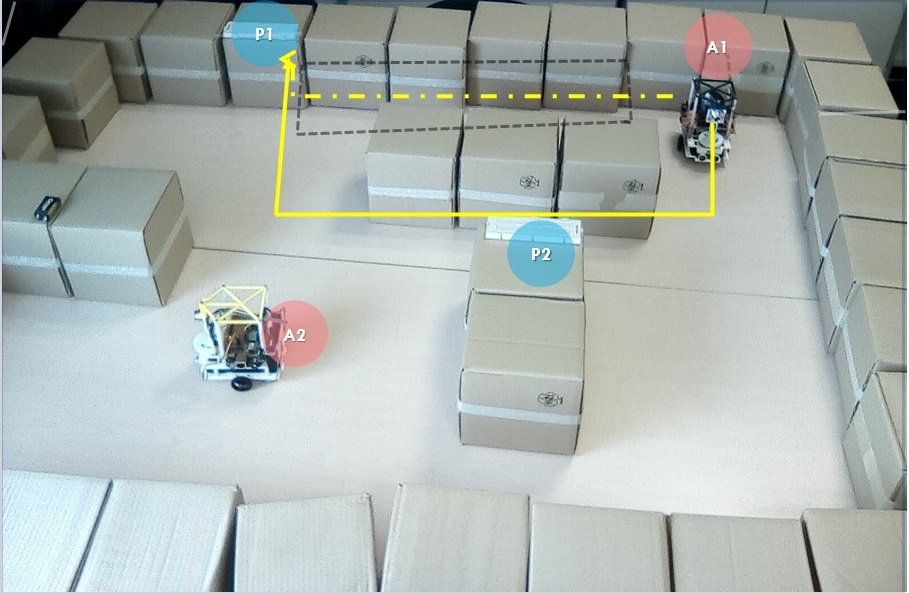}
\caption{An example}
\label{fig_exp}
\end{figure}
Figure~\ref{fig_exp} illustrates a scaled down automated internal transportation system, typically used for logistics in factories and executed by MRs. In this example, all MRs can execute only one task at a time. 
Let, at $t_i$, the path computed for $A$1 to carry some material to $P$1 is marked by the dotted line. Again, at time $t_j$ ($j>i$), $A$1 needs to carry same material to $P$1. But now, the battery capability of $A$1 has decreased due to execution of previous tasks and the condition of the given path has deteriorated (marked by dotted rectangle). Hence, more cost in terms of energy and time will be required to reach $P$1 at $t_j$ by $A$1. At this juncture, decision on routes can be improved using the real and current traversal cost to reach $P$1. This cost can be known directly by estimating the time to travel between two spots. The travel time is calculated considering the difference between the departure from one spot and reaching the next. The travel time is not dependent on the shape of the connection between two spots, rather it depends on the time taken to traverse between any two spots. These travel times can produce a different path than previous when used as a cost of traversal, as they include the dynamically changing factors. For example, the path marked by solid line is obtained considering the travel time as cost, which is a less cost consuming path to $P$1. 
\begin{figure*}[ht]
    \begin{subfigure}[b]{0.33\textwidth}
        \centering
        \includegraphics[scale=0.88]{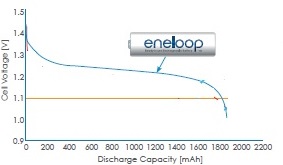}
        \caption{Discharge profile of batteries}
    \end{subfigure}%
    ~
    \begin{subfigure}[b]{0.33\textwidth}
        \includegraphics[scale=0.4]{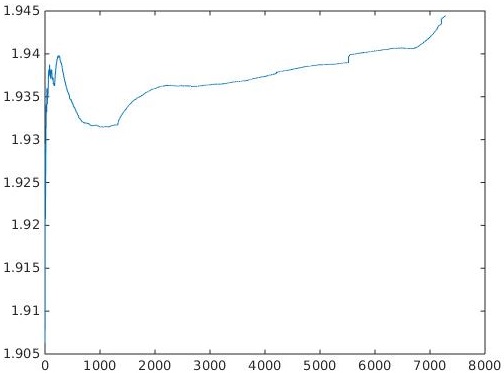}
        \caption{travel time with change of charge of batteries}
    \end{subfigure}
    ~
    \begin{subfigure}[b]{0.35\textwidth}
        \includegraphics[scale=0.555]{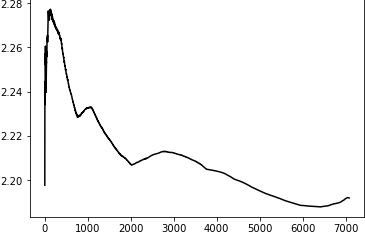}
        \caption{travel time with both the change of charge of batteries and floor}
    \end{subfigure}
    \caption{}
    \label{battvsTT}
\end{figure*}

In case of autonomous logistics, travel time arise locally at each MR due to action of actuators, wheels and other mechanical factors, but on the other they are significantly influenced by environmental factors like battery capacity (in case of battery powered MRs), conditions of the floor, conditions of material to be transported, performance and behavior of other AGVs, \textit{et~cetera}, as all or most of these factors determine the state of the robot at every instance of time. Influence of factors like friction forces of floor, slope, mechanical part can be corrected by local control on individual MR (lower levels), but factors like traffic condition, conditions of material, behavior of other MRs are beyond the scope of control by lower levels. Hence, considering cost coefficients at lower levels of actuation and control cannot make better control decisions. So these parameters are investigated at higher level to utilize them efficiently. These costs are time dependant and have sources of error from battery exhaustion, surface condition of shop floor, wear and tear of tyre, \textit{et~cetera}. Hence over the passage of time, the values will change. For example, Part (b) in Figure~\ref{battvsTT} plots the progressive mean of observed values of travel time for an $m$th edge first only with the change of state of charge of batteries and Part (c) with both the change of state of the charge of batteries and the floor condition from rough at the beginning to smooth. Part (a) of Figure~\ref{battvsTT} plots the cell voltage over time of Li-ion batteries.

The plot of progressive mean of travel time of (b) shows that the values increase first, then steadily decrease and then increase gradually till complete discharge. But the longer increase of values of travel time in (c) can be attributed to the rough floor, because at equal battery capacity in both cases, more energy is required to traverse in rough surface. Plot of the travel time of same arc in different conditions of floor demonstrate that travel time can reflect not only state of charge of batteries \cite{wafPragna} but also environmental conditions. 

In order to show how these variations in cost affect planning, standard Dijkstra's planning algorithm is modified to compute minimum cost paths. In fact, accurate and close-to-real estimated travel times can be used in any path finding algorithm. Dijkstra's algorithm is modified to use travel time as cost of edge, instead of heuristic cost based on distance. It is interesting to mention here that route planning is done in Tesla's new X 75D model cars according to battery need. The route planner proposes breaks of variable times to enable recharging while travellers enjoy recess in driving. Thus, states of battery and environment are considered for optimal route planning in these models. 
The total travelling costs of two categories of paths obtained by heuristically gathered cost and realistic travel time, irrespective of the route planning method, are calculated. Average total costs of path, obtained using static estimation of travel time (variation of edge travel cost over time is not considered) is roughly 5\% less that of paths, obtained by heuristics costs (Section~\ref{exp1}). However, travel times of edges (edge costs) vary along time and require to be predicted accordingly during path planning. A good estimation method to accurately predict travel times requires their histories, which can be collected progressively during MR operation. The estimation process start with mean of data obtained from legacy and real observations are obtained during the operation. Thus, the filtering method cannot generate the best estimates at initial few iterations. The estimates get improved over time. Real travel times are obtained by these estimations which can produced different paths than that of other costs like theoretical, heuristic and experimental. In fact, estimating traversal time by Kalman filtering shows that heuristic edge costs can underestimate total costs and, thus, can lead to non-minimal paths.
\subsection{Related works}
The two prominent problems of MRs are autonomous navigation and task scheduling. The autonomous navigation is addressed as a general problem of MR working in any unknown and dynamic environment, while task scheduling is a problem for MRs specifically operating in automated manufacturing units and warehouses. Usually, planning for navigation requires two different but complementary objectives, path planning and trajectory planning \cite{montiel2015path}. There are recent investigations to consider dynamic cost in path planning \cite{cho2017cost}, \cite{fazlollahtabar2018hybrid}, \cite{melo2015multi} yet cost is derived based on the distance between current node to next node or heuristics, not on dynamically changing conditions of environment and battery, though these factors are present on unknown terrain. 
The dynamic factors in automated factories are floor condition, state of battery, mechanical parts of robot while parts like racks, handlers, \textit{et~cetera} remain static mostly. This work addresses to consider these dynamic factors and study their effects in the planning for MR in automated manufacturing and logistics.

In case of proposals dealing with trajectory planning, dynamic cost based on time or energy is either derived out of motor dynamics \cite{kim2014online} or current pose and constant velocity \cite{kollmitz2015time}. In \cite{kollmitz2015time}, cost is not truly represented as it does not consider the change in battery states and environment which induces change in velocity. And in \cite{kim2014online}, the dynamics need to be changed for every new kind of robot model. In current work, travel time is considered as a cost which represents change in states of battery state and floor \cite{wafPragna} and can be derived similarly in any robot. 

Task scheduling, on the other hand, is addressed by introducing several constraints to each task like delivery time, location, transportation capacity of robots, \textit{et~cetera} \cite{veloso2015cobots}. Scheduling addresses to accomplish each task within the specified time taking into account all the constraints. In this context, this work proceeds one step further to estimate these necessary completion times for each task considering the state of charge of batteries and environmental conditions, so that minimum cost in terms of energy and time is expended to accomplish each task. In this work, a new method is proposed to find costs for performing tasks like traversing between spots in order to decide optimally. Here, path planning is considered as an example of planning It is done for a single MR and costs incurred in traversing are predicted by estimating travel times between the spots. Few works on general problem of road-map generation for MRS in logistics have also considered cost as dynamic \cite{digani2015ensemble}, however, it is based on the Euclidean distance.

To best of author's knowledge, travel time is not considered as a dynamic cost for planning in MRS. 
\subsection{Contribution}
The contribution of this paper is twofold. Firstly, a new method to estimate realistic transportation costs in automated logistics is proposed. Secondly, the dynamic route or task planning for an MR has been improved using these estimated values. 
\subsection{Organization of the paper}
The next section Section~\ref{probdefn} formulates the problem in the light of an internal logistic system with path traversal as a task. Section~\ref{proto} explains the prototype platform and other details for the system used to conduct experiments. Two experiments and their results are elaborated in Sections~\ref{exp1} and ~\ref{exp2} with Algorithm~\ref{staticCost} elaborating on the proposed approach which incorporates modification over Dijkstra's algorithm. Section~\ref{last}
concludes with discussions and future directions of investigation.
\section{Problem Formulation}
\label{probdefn}
In this work, the focus is on one MR. Also, traversing a path is considered as a task (Section~\ref{intro}). Path traversal is a sequence of traversal from one spot to another, expressed as nodes. A path $P$ for a robot is usually defined as,
\begin{equation}
\label{introP}
\begin{aligned}
P = \langle(n_a, n_b), (n_b,n_c), (n_c,n_d), (n_d,n_e), ...........)\rangle
\end{aligned}
\end{equation}
where $n_p$ is any node. $P$ can be also expressed in terms of connecting edges as 
\begin{equation}
\label{edgewidP}
\begin{aligned}
P = \langle a_{a,b}, a_{b,c}, a_{c,d}, a_{d,e}, ...........\rangle
\end{aligned}
\end{equation}
where $a_{p,q}$ is any edge.

Travel time determines the cost of such traversal tasks.
The total travel time of these segments determines the total travelling cost of a path.
As explained in Section~\ref{intro}, each edge in the floor map is associated with some cost in terms of energy exhaustion and others. Hence, traversing a defined path incurs several travel costs for all edges in the path. Thus time to traverse an edge by a MR can be conceptualized as its cost coefficients. 
Let $X_{p,q}(e,f)$ denote edge cost from $n_p$ to $n_q$, where $e$ denotes dependency for state of charge of batteries and $f$ denotes dependency for frictional force of the floor. For simplicity, $X_{p,q}(e,f)$ will be denoted as $X$ from now onwards. Now, the total cost of traversing $P$ can be written in a form $C_p$ as, 
\begin{equation}
\label{PwithX}
\begin{aligned}
C_p = \langle(X_{a,b}(e,f), X_{b,c}(e,f), X_{c,d}(e,f),...,...)\rangle
\end{aligned}
\end{equation}
In equation~\ref{PwithX}, it is shown that a total path cost is dependent on all edge costs and each edge cost is denoted by its travel time. Thus, $X$s denote general travel time of any edge. Also, travel time of any edge $X$ depends on all the previous edges the robot has already traversed. The reason being the discharge of batteries and (or not) possible change of environment. Thus, travel time $X$ becomes a function of $k$ as $X$($k$) where, $k$ = number of time a MR has performed the task of traversing any edge. 

Hence, $X$($k$) is estimated with respect to increase in $k$ for any edge and used as weight of edge to compute path. The estimated value of $X$($k$+1) depends only on $X$($k$) and the observation of $X$ at ($k$+1). These experiments and results are explained in Section~\ref{exp1}. Observations of all possible $X$($k$) for all possible $k$ need to be made for this above estimation for a single MR. 

However, this is not only cumbersome but also impractical to gather such huge amount of observations. This estimation is static as $X$ is estimated without considering its variation with the total elapse of time from start of system. 
The static estimation approach is progressed to a different model. A window of current $X$ and a fixed number (let $j$) of values of previous $X$s are used to form a state vector. The previous values of $X$s are the travel times of those edges which are already considered to form the path. Also, state vector contains an exploration variable $\xi$($k$). A fixed window of values of $\xi$($k$) of same size $j$ is considered in the state vector. Thus, the state vector contains both parts, the change of $X$ over time and the $X$s of previous edges. The state vector (let $s$) is estimated on every $k$ and $s$($k$+1) is formed. $X$($k$+1) is one element of $s$($k$+1). Thus, $X$($k$+1) is formed for every $k$. Here, the current value of $X$ is estimated depending on the previous $X$s i.e.-travel times of edges along with the variation of exploration of $X$ due to elapse of time. Thus, $X$ values are dynamically estimated considering its variation over elapse of time. Moreover, the model is allowed to gather the possible values of $X$ itself from the beginning of first call of path planning and use these values to estimate current value. Observations of all possible $X$($k$) for all possible $k$ are not needed in the latter. This experiment is elaborated in Section~\ref{exp2}.
\section{Prototyped Internal Transportation System}
\label{proto}
\begin{figure}[h]
\centering
\includegraphics[width=0.45\textwidth]{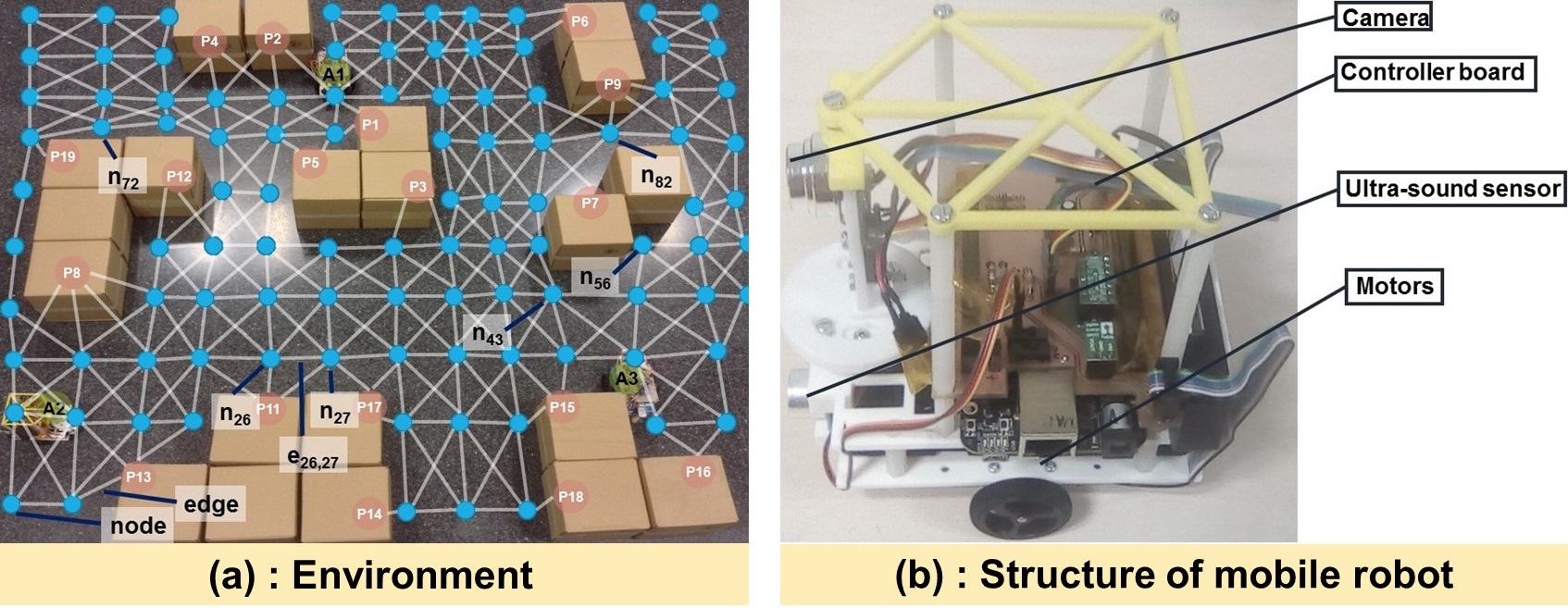}
\caption{Scale downed prototype platform}
\label{envrn&map}
\end{figure}

\begin{figure*}
\includegraphics[scale = 0.378]{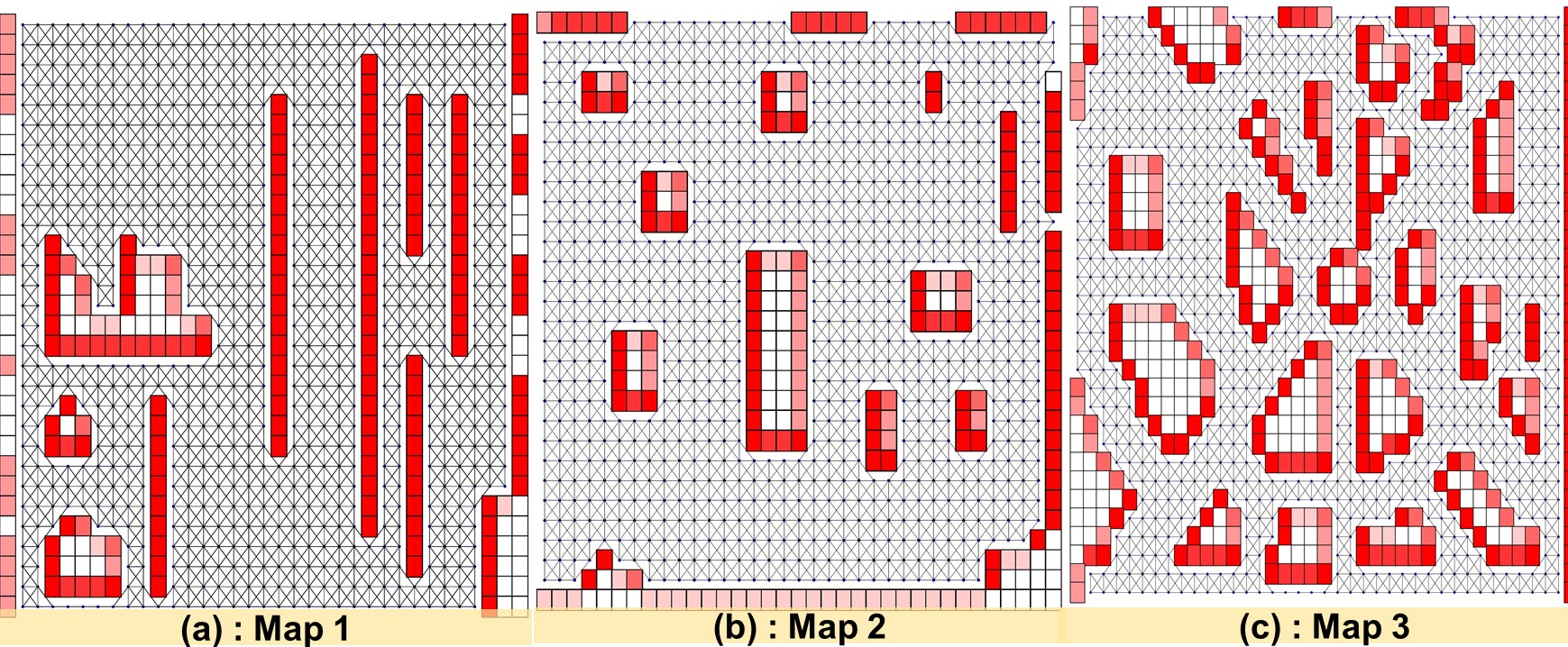}
\caption{Three representative topological maps}
\label{3maps}
\end{figure*}

A prototype scaled down internal transportation system is developed with all essential constituting parts like MRs, tasks, controller architecture and the environment adhering to minute details. The floor is described by means of a topology map $\mathcal{G} = \{\vee,\varepsilon\}$, where each port and bifurcation point corresponds to some node $n_r$ $\in \vee$ and each link between two nodes forms an edge $a_{e,f}\in \varepsilon$. Part (a) of Figure~\ref{envrn&map} depicts a portion of the whole prototype, 
where, notation like $n_{26}$ designates a node and $a_{26,27}$ a edge.  
Topology maps are generated taking reference from the grid map generated by results of Simultaneous Localisation and Mapping (SLAM) in \cite{beinschob2015graph} based on a simple assumption, that each free cell in the grid map corresponds to a node in the graph. The SLAM from \cite{beinschob2015graph} produces mapping and localization of the Coca-Cola Bilbao plant with good accuracy and thus can be treated as a map of the real plant. 
A selected representative portion from each of the three sections of of Coca-Cola Iberian Partners in Bilbao, Spain are extracted to form three topological maps. They are provided in Figure~\ref{3maps}. Part (a) of Figure~\ref{3maps} illustrates Map~1 which is a representative of winding racks in the warehouse facility, Part (b) shows Map~2 which represents randomly placed racks and Part (c) shows Map~3 which represents racks organized in a hub. 

The scaled robots are built controller, ultrasound sensor and a camera, as illustrated in Part (b) of Figure~\ref{envrn&map}. The DC servo motors drive the wheels of the MR and Li-ion batteries energize them. Each MR has its individual controller in decentralized architecture \cite{ribas2013agent, ribas2009approach}. 

Travel times for three different length of arcs in all three maps and four different conditions of surface are recorded till complete exhaustion of batteries. This generates observation data for all possible $X$s for all $k$s. This cumbersome process of acquiring the observations is mitigated by a non-linear functional model of $X$ which allows to gather information about $X$ for different arcs in the map gradually with time. After start of computing a path, the real travel time of edges are recorded when the MR actually traverses it. This travel times of edges are used as the observation values for the next call of path planning. Thus observation values of travel times of each edge is grown during run-time. 
\section{Experiment I: Using static estimates of travel times in route planning}
\label{exp1}
\subsection{Procedure}
The state-space model provided in equations~\ref{state_eqn} and \ref{obs_eqn} is used to estimate travel times. From now $X_{p,q}$ will be written as $X$ for simplicity,  
\begin{align} 
\label{state_eqn}&X(k)= X(k-1) + \omega(k)\\
\label{obs_eqn}&Y(k)= X(k) + \eta(k)
\end{align}
The state vector in equation~\ref{state_eqn} is a single variable $X$ depending on the number of edges already found in the path, $k$ (Section~\ref{probdefn}). Hence, $X$ is estimated over and over again for different connecting edges of every new exploring node. $Y$($k$) in equation~\ref{obs_eqn} is the observation variable for $X$. This model involves two error terms $\omega$($k$) and $\eta$($k$) which are independent and normally distributed. According to equations~\ref{state_eqn} and \ref{obs_eqn}, $X$($k$) depends only on the travel time of edge between current node and its predecessor i.e.-$X$($k$-1), though in reality, it depends on $X$s for all the previous edges in the path and its own variation over the time. This estimation process is static as it does not consider the change of $X$ for the total elapse of time since start of system (Section~\ref{probdefn}). 
Equations~\ref{KF_state1} and \ref{KF_state2} are obtained after applying Kalman Filtering method \cite{kfRibeiro} on equations~\ref{state_eqn} and \ref{obs_eqn}.
\begin{equation}
\begin{aligned}
\label{KF_state1}&\hat{X}^{-}(k)= \hat{X}(k-1)  \\
&P^{-}(k)= P(k-1) + \sigma^2_\omega 
\end{aligned}
\end{equation}
\begin{align}
\label{KF_state2}
&K(k)=P^{-}(k)\big/[P^{-}(k)+\sigma^2_\eta] \nonumber \\
&P(k)=P^{-}(k)-[{P^{-}(k)}^2\big/[P^{-}(k)+\sigma^2_\eta]]\\ \nonumber
&\hat{X}(k)=\hat{X}^{-}(k)+[P^{-}(k)\big/P^{-}(k) 
+\sigma^2_\eta]*\omega(k)\\ \nonumber
&\text{where,}\qquad  \omega(k) = [Y(k)-\hat{X}^{-}(k)] 
\end{align}
$\hat{X^{-}}(k)$ produces the apriori value of $X$ and $P^{-}$ produces the associated covariance, $\sigma^2_\omega$ being the covariance of process noise $\omega$($k$). $\hat{X}$($k$) provides the predicted estimate of $X$($k$), as $\hat{X^{-}}$($k$) is corrected in equation~\ref{KF_state2} with the help of Kalman Gain $K$($k$). $P^{-}$($k$) provides the associated covariance matrix, $\sigma^2_\eta$ being the covariance of the observation noise $\eta$($k$).

\begin{figure}[t]
\centering
\includegraphics[scale = 0.35]{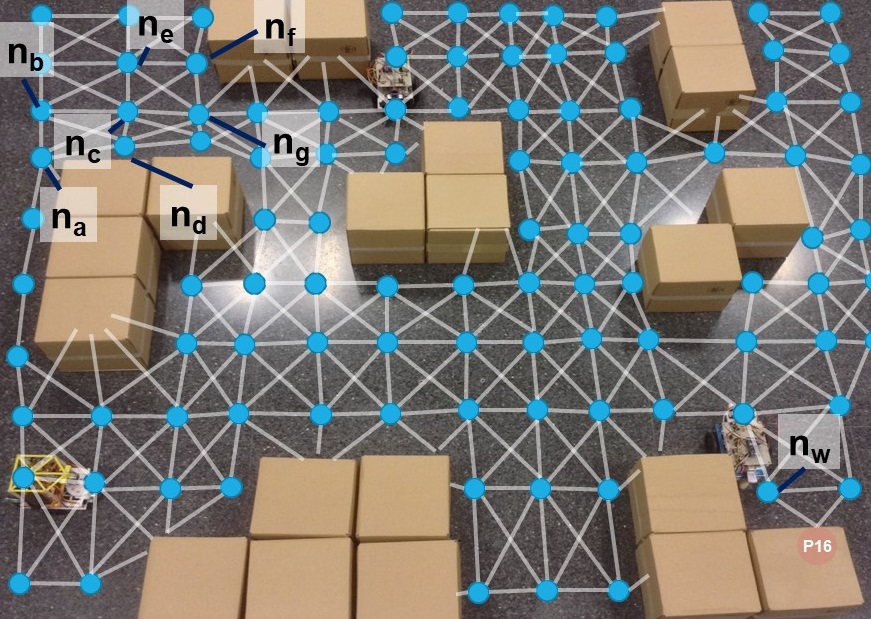}
\caption{Sample run of route computation}
\label{samplerouteplan}
\end{figure}

For example, a sample route computation is illustrated in Figure~\ref{samplerouteplan}. Let $n_a$ be source and $n_w$ destination at $P$16. 
So, path computation using Dijkstra's algorithm starts at $n_a$ with its neighbors $n_b$, $n_c$ and $n_d$. So, $X_{a,b}$, $X_{a,c}$, $X_{a,d}$ are required to be estimated at $k$, when $k$ is 1 as this will be first edge being traversed. 
\begin{align}
\label{xinitial}
&\hat{X}(0)= E[X(0)]\\
\label{Pinitial}
&P(0)= E[(X(0)-E[X(0))(X(0)-E[X(0))^T]
\end{align}
We use equation~\ref{KF_state1} to obtain $\hat{X^{-}}$(1) for $X_{a,b}$, $X_{a,c}$, $X_{a,d}$ separately depending on $X$(0) using equation~\ref{xinitial}. Similarly, we get separate $P^{-}$(1) using equation~\ref{Pinitial}. Next, we obtain $\hat{X}$(1) (estimate) and $P$(1) for $X_{a,b}$, $X_{a,c}$, $X_{a,d}$ using equation~\ref{KF_state2}. 
Comparison of estimated values of $X_{a,b}$, $X_{a,c}$, $X_{a,d}$ will provide the least cost of traversing from $n_a$ to any of its neighbor. Let, the least cost edge be $a_{a,c}$. So $n_a$ will become the predecessor of $n_c$, i.e.-to reach $n_c$, the edge should come from $n_a$. When $n_c$ will be explored, the value for $k$ is 2 as $n_c$ has 1 predecessor. The next least cost edge from $n_c$ in the path is required to be known. Thus, $X_{c,e}$, $X_{c,f}$, $X_{c,g}$ needs to be estimated. Thus, observation $Y$($k$) of $X$ at current $k$ is required to estimate $X$. Thus observation values for travel costs of all possible $X$s for all possible $k$s were collected. This above process is explained in Algorithm~\ref{staticCost}. 
This static experiment is conducted to verify that weights of edges can be estimated online during exploration of Dijkstra's algorithm using a state-space model. Also, it is verified that the estimated values of $X$ are correct and real through this experiment, as the values can be compared to real observations. 
\subsection{Results} 
\label{resexp1}
\begin{figure*}
\includegraphics[scale = 0.20]{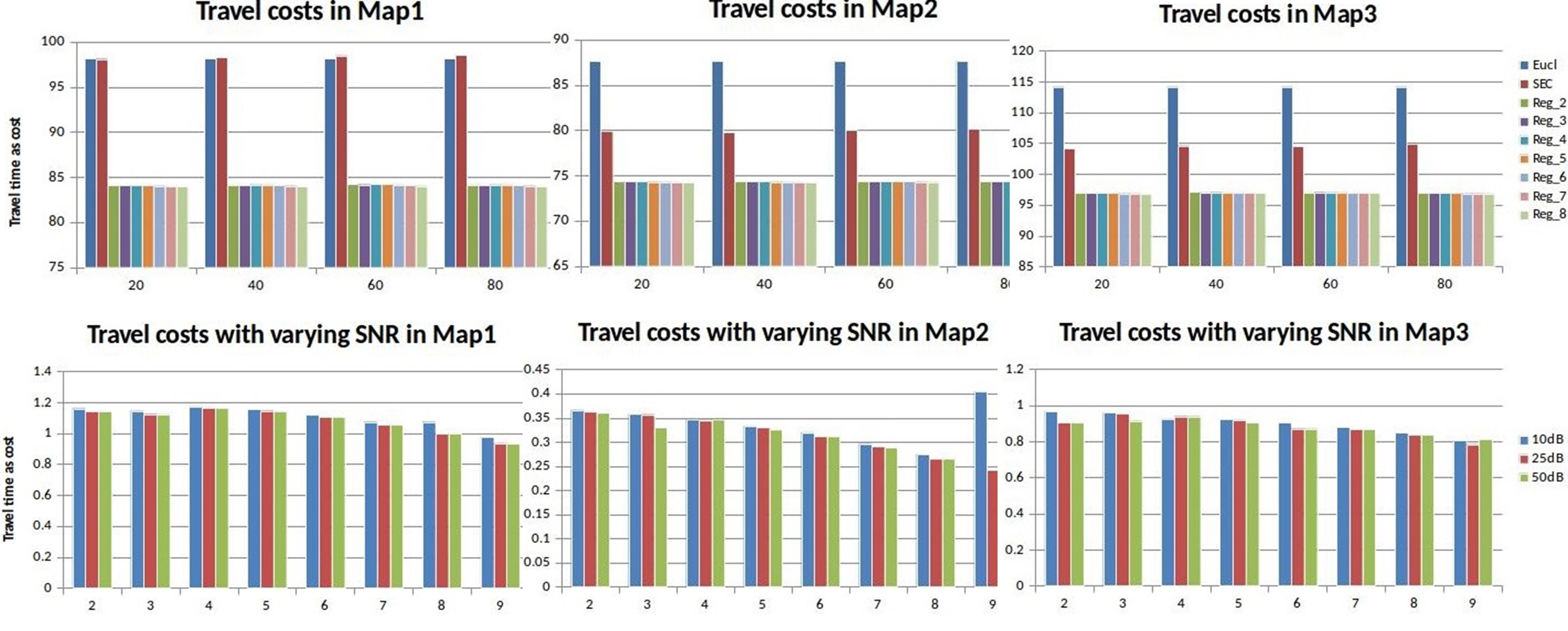}
\caption{Comparison cost}
\label{compareCost}
\end{figure*}

Paths are computed repeatedly for 20, 40, 60 and 80 times in each topological graph (Figure~\ref{3maps}) using both original Dijkstra's algorithm using Euclidean distance based heuristic weights of edges and the modified one (Algorithm~\ref{staticCost}). The choices of source and destination are fed from the decided list of sources and destinations for each call of route computation. Total path costs obtained using heuristic weights are compared with paths obtained using Algorithm~\ref{staticCost}. The vertical bars of \textit{Eucl} and \textit{SEC} in Figure~\ref{compareCost} represent the average total path costs for heuristic cost based routes and static estimates based routes respectively. Vertical bar \textit{Eucl} shows that average total path costs never change with increase in number of repetitive calls, as heuristic weights do not change over time and does not reflect the true cost of traversal.
\begin{algorithm}
 \caption{Using static estimation of travel time in Dijkstra's algorithm}
 \label{staticCost}
 \begin{algorithmic}[1]
    \Function{initialise\_single\_source}{$\vee,s$}
    \Comment{Where $\vee$ - list of nodes, s - source, returns $d[v]$ - atribute for each node, $\pi[v]$ - predecessor for each node}
 \For{each $x_i \in V$}
    \State $\pi[x_i] = infinity$
    \State $d[x_i] = NIL$
  \EndFor
    
 \EndFunction
  \Function{find\_prev}{$(u,s)$}
    \Comment{input: $u$-current node,$s$-source node, returns: $prev\vee$-predecessor of $u$, $noPred$ -number of predecessors till $s$}
       \State $prev\vee$ = compute predecessor of $u$
       \State $noPred$ = count of predecessors till $s$
        \EndFunction
    \Function{KF}{$(pW,k,Y(k))$}\Comment{input: $pW$-value of travel time at $k$ -1, $k$-instance for estimation, $Y$- observation variable, Returns: $\hat{X}$($k$)-travel cost from $u$ to $v$}        
    \State Apply KF on state-space model to obtain $\hat{X}$($k$)
    \EndFunction
    \Function{find\_cost}{$u,v,k,pW,Y(k)$}
    \Comment{Input: $u$-current node, $v$- neighbor node, $k$- instance of estimation,$pW$ - cost between prev$u$ and $u$, $Y$($k$) - observation of travel time between $u$ and $v$
    ,Returns:$w$- estimated travel\_time (cost) from $u$ to $v$}
        \State $w$ = $KF$($pW$,$k$,$Y$($k$))
     \EndFunction
     \Function{Relax}{$u,v,w$}
    \Comment{Inputs: $u$-current node, $v$- neighbor node, $w$- estimated travel\_time (cost) from $u$ to $v$, Returns: $d$[$v$]-attribute for each each node, $\pi$[$v$]-predecessor of each node}
    \If{$d$[$v$] $> d$[$u$] + $w$($u,v$)}
      \State $d[v] = d[u] + w(u,v)$
      \State $\pi[v] = u$
      \EndIf    
    \EndFunction
\Function{Main}{$\vee, \varepsilon, Y, s$}
    \Comment{Inputs: $\vee$-list of nodes, $\varepsilon$-list of edges, $s$-source node, $Y$-observation matrix,Returns: $\pi$[$v$]-predecessor of each node, $w$-edge weight matrix}
    \State $P$ := NIL
    \State $Q$ := queue($\vee$)
    \State $k$ := 0
   \State $p\varepsilon$[$s$] = 0
    \State $w$[$p\varepsilon$[$s$], $s$] = 0
    $initialise\_single\_source$($\vee,\varepsilon,s$)
    \While{$Q != $0}
       $u$ := Extract min-priority queue($Q$)
       $p\vee$[$u$], $npred$ = $find\_prev$($u,s$)
       $k$ = $npred$+1
       $pW$ = $w$[$p\varepsilon$[$u$], $u$]
       $P$ := $P$ $\bigcup u$
       \For{each $v \in Adj$[$u$]}
         \State $w$[$u$,$v$] = $find\_cost$($u$,$v$,$k$, $pW$,$Y$($k$))
         \State $relax$($u$,$v$,$w$)
         \EndFor
       \EndWhile
    \EndFunction
 \end{algorithmic}
\end{algorithm}


Vertical bar \textit{SEC} shows that average total path costs obtained by Algorithm~\ref{staticCost} is 5\% less in case of Map~2 and Map~3 and 2\% less in Map~1 than that of heuristic cost based Dijkstra's algorithm. Average total path costs increases with number of repetitions as shown by vertical bar \textit{SEC}, as duration of performance increases with increase of repetitions. This happens due to the dependency of current edge cost on previous edge cost (equations~\ref{state_eqn} and \ref{obs_eqn}). But, this variation does not truly reflect the variation of travel time due to time-varying factors.
\section{Experiment II: Using dynamic estimates of travel times in route planning}
\label{exp2}
\subsection{Procedure}
The bi-linear model \cite{priestley1988}, provided in equation~\ref{bilinear} is used to model the change of travel costs depending upon all the previous travel costs. $X$ is formed as a function of its past histories over $k$, considering the progressive change $\xi$ with respect to $k$. After start of computing a path, the real travel time of edges are recorded when the MR actually traverses it. This travel times of edges are used as the observation values for the next call of path planning. Thus observation values of travel times of each edge is grown during run-time.  
\begin{align*}
X(k)+a_1X(k-1)+.....+a_jX(k-j)
= \xi_k+b_1\xi(k-1)\\ \nonumber
+...+b_l\xi(k-l)
+\sum\sum c_{rz}\xi(k-r)X(k-z)
\end{align*}
The double summation factor over $X$ and $\xi$ in the above equation 
provides the nonlinear variation of $X$ due to state of batteries and changes in environment. The state space form of the bi-linear model is given in equations~\ref{bStateEqn} and \ref{bObsEqn}.
In equation~\ref{bStateEqn}, the state vector $s$($k$) is of the form $(1,\xi(k-l+1),....,\xi(k),X(k-j+1),......,X(k))^T$. 
Here, $j$ and $l$ denote number of previously estimated $X$s and previous innovations of $X$ respectively. The term $regression\_no$ denotes the values of $j$ and $l$ and is chosen as a design parameter. The $regression\_no$ is increased from 2 to 9 and the effects on total edge travel cost of paths is demonstrated in Section~\ref{exp2results}. 
\begin{align}
    \label{bStateEqn}
&s(k) = F(s(k-1))s(k-1) + V\xi(k)+G\omega(k-1)
\end{align}
\begin{align}
\label{bObsEqn}
&Y(k) = Hs(k-1) + \xi(k)+ \eta(k)
\end{align}
The state transition matrix $F$ in the equation~\ref{bStateEqn} has the form of
\[
F =
\adjustbox{width=0.7\linewidth}{
$\begin{bmatrix}
    1\mspace{18.0mu}0\mspace{18.0mu} 0\mspace{18.0mu}\dots\mspace{18.0mu}0\mspace{18.0mu}\vdots\mspace{18.0mu}0\mspace{18.0mu} 0\mspace{18.0mu}\dots\mspace{18.0mu} 0\mspace{18.0mu}0\\
    0\mspace{18.0mu}0\mspace{18.0mu} 1\mspace{18.0mu}\dots\mspace{18.0mu}0\mspace{18.0mu}\vdots\mspace{18.0mu}0\mspace{18.0mu} 0\mspace{18.0mu}\dots\mspace{18.0mu} 0\mspace{18.0mu}0\\
    0\mspace{18.0mu}0\mspace{18.0mu} 0\mspace{18.0mu}\dots\mspace{18.0mu}1\mspace{18.0mu}\vdots\mspace{18.0mu}0\mspace{18.0mu} 0\mspace{18.0mu}\dots\mspace{18.0mu} 0\mspace{18.0mu}0\\
    0\mspace{18.0mu}0\mspace{18.0mu} 0\mspace{18.0mu}\dots\mspace{18.0mu}0\mspace{18.0mu}\vdots\mspace{18.0mu}0\mspace{18.0mu} 0\mspace{18.0mu}\dots\mspace{18.0mu} 0\mspace{18.0mu}0\\
    \vdots\mspace{18.0mu}\vdots\mspace{18.0mu}\vdots\mspace{18.0mu}\dots\mspace{18.0mu} \vdots\mspace{18.0mu}\vdots\mspace{18.0mu}\vdots\mspace{18.0mu} \vdots \mspace{18.0mu}\dots\mspace{18.0mu} \vdots\mspace{18.0mu} \vdots\\
    0\mspace{18.0mu}0\mspace{18.0mu} 0\mspace{18.0mu}\dots\mspace{18.0mu}0\mspace{18.0mu}\vdots\mspace{18.0mu}0\mspace{18.0mu} 1\mspace{18.0mu}\dots\mspace{18.0mu} 0\mspace{18.0mu}0\\
    0\mspace{18.0mu}0\mspace{18.0mu} 0\mspace{18.0mu}\dots\mspace{18.0mu}0\mspace{18.0mu}\vdots\mspace{18.0mu}0\mspace{18.0mu}0 \mspace{18.0mu}1\mspace{18.0mu}\dots\mspace{18.0mu}0\\
    0\mspace{18.0mu}0\mspace{18.0mu} 0\mspace{18.0mu}\dots\mspace{18.0mu}0\mspace{18.0mu}\vdots\mspace{18.0mu}0\mspace{18.0mu} 0\mspace{18.0mu}0\mspace{18.0mu}\dots\mspace{18.0mu} 1\\
    \mu\mspace{18.0mu}\psi_l\mspace{18.0mu}\psi_{l-1}\mspace{18.0mu}\dots\psi_1\vdots-\phi_j-\phi_{j-1}\dots-\phi_1
\end{bmatrix}$}
\]
The number of rows of $F$ is given by (2*$regression\_no$ + 1). 

The $\psi$ terms in $F$ are denoted as in equation~\ref{psiterms}
\begin{equation}
\label{psiterms}
\begin{aligned}
\psi_l =b_l+ \sum_{i=1}^{l} c_{li}X(k-i)\\
\end{aligned}
\end{equation}
All the $\phi$ terms in $F$ are constants. The term $\mu$ is the average value of $X$ till $k$. 
Also, the matrix $V$ in \ref{bStateEqn} is denoted as 
\[
V = 
\begin{bmatrix}
    0 & 0 & 0 & \dots & 1 & \vdots & 0 & 0 & \dots & 1
\end{bmatrix}
\]
The number of rows of $V$ is again given by (2*$regression\_no$ + 1). The matrix $H$ in \ref{bObsEqn} is denoted as 
\[
H = 
\begin{bmatrix}
    0 & 0 & 0 & \dots & 0 & \vdots & 0 & 0 & \dots & 1
\end{bmatrix}
\]
Kalman filtering is applied on the state-space model (equations~\ref{bStateEqn} and \ref{bObsEqn}) resulting in equations~\ref{KF1dynamic} and \ref{KF2dynamic} to estimate $s$ repeatedly to obtain $X$ for the connecting edges at each node to compute path using Dijkstra's algorithm.  
\begin{align}
\label{KF1dynamic}
&\hat{s^{-}}(k) = F(s(k-1))s(k-1) + V\xi(k)+G\omega(k-1)\\ 
&\hat{P^{-}}(k) = F(s(k))P(k-1)F^T(s(k-1))+ Q(k-1) 
\end{align}
In equation~\ref{KF1dynamic}, $\hat{s^{-}}$($k$) provides the apriori estimate of $s$. $\hat{P^{-}}$ provides the associated covariance matrix where $Q$($k$-1) provides the covariance for the process noise $\omega$($k$-1).
\begin{align}
\label{KF2dynamic}
&K(k)= \hat{P^{-}}(k)H^T[H\hat{P^{-}(k)}H^T+R(k)]\\ \nonumber
&\hat{s}(k) = \hat{s^{-}}(k)+K(k)[Y(k)-H\hat{s^{-}}(k)]\\ \nonumber
&P(k) = [I - (K(k))H]\hat{P^{-}}(k) 
\end{align}
In equation~\ref{KF2dynamic}, $K$($k$) is the Kalman gain, $R$($k$) being the covariance of observation noise $\eta$($k$). $\hat{s}$($k$) provides the estimated state vector $s$ at $k$. 
\begin{align}
\label{sinitial}
&\hat{s}(0)= E[s(0)]\\ 
\label{P2initial}
&P(0)= E[(s(0)-E[s(0))(s(0)-E[s(0))^T]
\end{align}
In Figure~\ref{samplerouteplan}, the path computation starts at $n_a$. Let the values of $j$ and $l$ are equal which is 2. At start, $k$ is 1. Now $s$ cannot be formed as minimum 2 previous travel costs are needed. Exploration proceeds with average travel cost for the edges. When $n_c$ needs to be explored, value of $k$ becomes 2 as one travel cost has been known connecting $n_c$ to its predecessor $n_a$. $s$ can be now formed as $X$(1) is known. Again, $n_a$ is the source and so $X$(0) is 0. $\xi$ is assumed to be $N$(0.1,0.1). At $k$ =2, $s$(1) takes the form $(1, \xi(0),\xi(1), X(0), X(1))^T$. Equation~\ref{KF1dynamic} and \ref{KF2dynamic} are used to estimate $s$(2) separately for all edges arising out of $n_c$ to obtain $X$ for each edge. From equations~\ref{sinitial} and \ref{P2initial}, $s$(0) and $P$(0) can be obtained. Let at $n_g$, $k$ = 4. Hence, $X$(3) will be travel cost from $n_e$ (predecessor of $n_g$) to $n_g$, $X$(2) will be travel cost from $n_c$ (predecessor of $n_e$) to $n_e$, $X$(1) will be travel cost from $n_a$ (predecessor of $n_c$) to $n_c$. Thus, $s$(3) = $(1,\xi(2),\xi(3), X(2), X(3))^T$ and $s$(4) = $(1,\xi(3),\xi(4), X(3), X(4))^T$ needs to be computed. This approach is different from Algorithm~\ref{staticCost} in the way the $X$ is estimated. 
\subsection{Results}
\label{exp2results}
The process of path computation is exactly similar to previous experiment. Only difference is in the estimation procedure of $X$, which is based on the bil-inear state space model. The $b$ and $c$ are chosen as normal distribution. Along with the repetitions of path computations, the value $\phi$, mean and covariance of $b$ and $c$ are increased from -0.4 to 0.4 and from -0.2 to 0.2 respectively. Negative values of $\phi$ produced too high estimates while values greater than 0.2 produced negative estimates. Similarly, mean and covariance values less than 0.1 produce high estimates and more than 0.1 produce negative estimates.
Thus, 0.2 is found as the suitable value of $\phi$ and $N$(0.1,0.1) suits for both $b$ and $c$. Also, the \textbf{$regression\_no$} from 2 to 9 for each of 20, 40, 60 and 80 repetitive computation and average total path costs obtained on each case are plotted in Figure~\ref{compareCost}. The vertical bars marked from $Reg_2$ to $Reg_9$ represent the average total path costs for dynamic estimates based routes, which shows that they are 15\% less on average than heuristic euclidean cost for all three maps in each set of repetitions. This difference is increased with the increase of \textit{$regression\_no$}, though the rate of increase is low, as the change of $X$ itself is not broadly spread with standard deviation of 0.219 on average. The average total path cost increases with increase in number of repetitions as edge travel cost increases with elapse of time. The observation $Y$($k$) developed during run-time is considered as signal and the values of $\omega$ are modified to increased the Signal-to-Noise Ratio (SNR) from 10dB to 50 dB along with the repetitions of path planning. The vertical bars marked 10dB, 25dB and 50dB in Figure~\ref{compareCost} plots the average path costs obtained by changing the SNR for each \textit{$regression\_no$}. which shows that with the increase of SNR, the average travel cost decreases.
\subsection{Path comparison}
\begin{figure*}
\includegraphics[scale = 0.36]{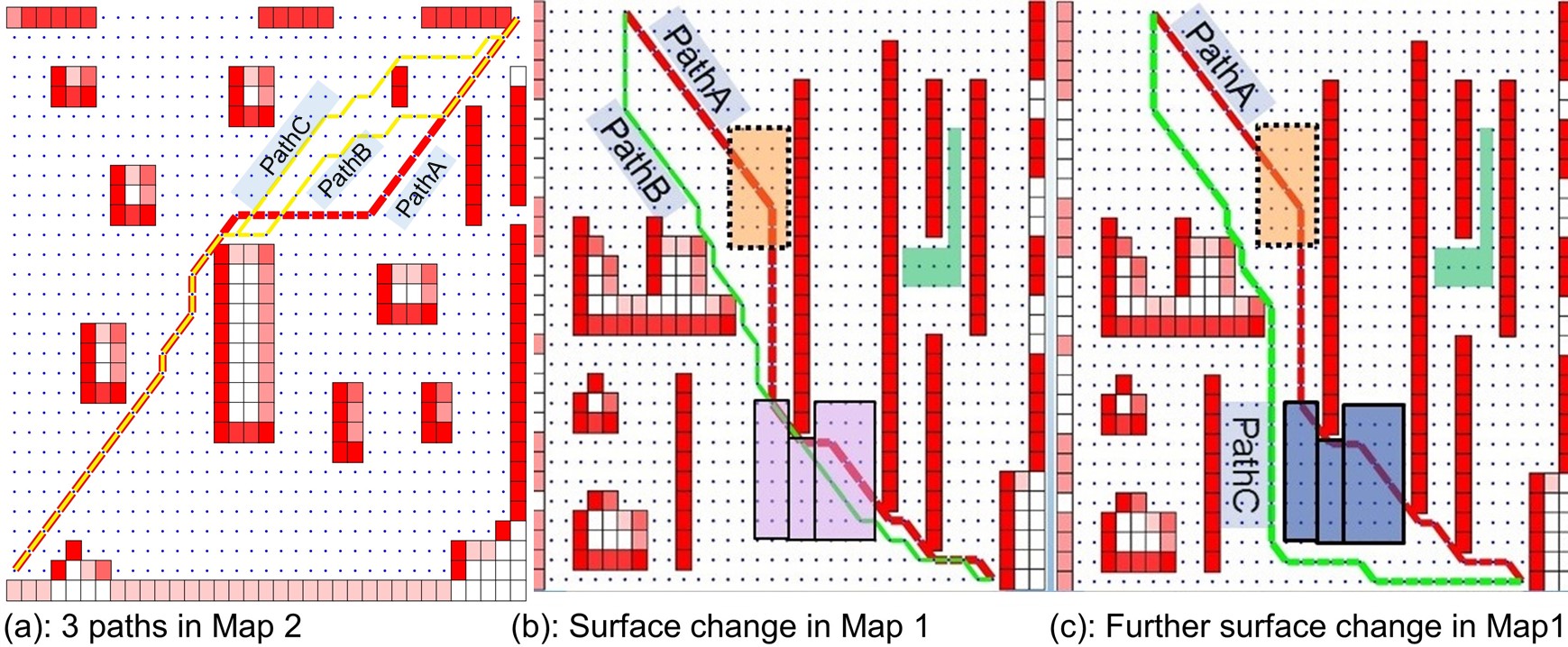}
\caption{Paths in different conditions}
\label{3paths}
\end{figure*}

Part (a) of Figure~\ref{3paths} plots 3 single paths Path$A$, Path$B$ and Path$C$ obtained from Dijsktra's algorithm based on heuristic costs, statically estimated and dynamically estimated edge travel costs respectively for the same pair of source and destination nodes in Map~2 including only the variation induced by discharge of batteries. 
Here, $P_{cA}$, $P_{cB}$ and $P_{cC}$ stands for the general $P_c$ vector explained in Section~\ref{probdefn} for Path$A$, Path$B$ and Path$C$ respectively. $P_{cA}$, $P_{cB}$ and $P_{cC}$ have many common elements, despite having different elements. Thus, the total travel cost in these 3 paths are different. After obtaining the total travel costs of Path$A$, Path$B$ and Path$C$, it can be stated that, 
\begin{align}
    \sum P_{cB} < \sum P_{cA} by 5\% \nonumber and 
    \sum P_{cC} < \sum P_{cA} by 15\% \nonumber
\end{align}
This also establishes the proposal that heuristics based path planning can underestimate real edge travelling costs and lead to expensive paths. 
Path$A$ and Path$C$ in (b) and (c) of Figure~\ref{3paths} are obtained in Map~1 by heuristic based edge weights and dynamically estimated edge travel costs respectively, when floor condition is changed in dotted line zone to moderately rough and solid line zone to lightly rough after 20 calls for route computation. Path$A$ in both (b) and (c) contains edges in both rough zones in the floor, while Path$C$ in (b) clearly avoids the zone with moderate roughness, though having few edges in the lightly rough zone. This happens because Dijkstra's algorithm finds that cost incurred in traversing the lightly rough zone to be less than that of the additional edges required to avoid the zone. This proves that modification of Dijkstra's algorithm using dynamically estimated travel cost does not disrupt the computational robustness of the algorithm. Also, when the lightly rough zone is made heavily rough, Path$C$ deviates to other direction and adding more nodes. Thus again, estimated travel times of edges help Dijkstra's algorithm to find a cost effective path. 
\subsection{Real cost saving for paths}
In (b) of Figure~\ref{3paths}, there are total 12 edges from the 2 rough zones comprised in Path$A$. The path cost of Path$A$ obtained using heuristic weights is not the correct one as travel costs of each of these 12 edges are more than assumed. Let, a variable $\delta$ accounts for the additional edge costs in each edge. Path cost of Path$A$ is obtained as 98.210 from results, but in reality path cost of Path$A$ should be (98.210 +12*$\delta$). The value of $\delta$ can never be zero as changes in environment ans batteries will always be present. When more zones will have changed floor conditions, more edges will have increased edge cost. So, the coefficient of $\delta$ will increase and also the true value of travel cost of paths. Thus, the difference between travel costs of paths obtained by heuristic cost and estimated travel time will always increase with the increase of hostility in the environment.
\section{Discussion and Conclusion}
\label{last}
The travel times of edges are identified as one of the cost coefficients in internal automated logistics. A formulation is devised to estimate travel times online during path computation considering its time-varying components. Moreover, suitable observations for travel time are recorded in scenarios with analogy to real factory in a scaled platform developed in the laboratory. They are instrumental for feeding into estimation algorithms to estimate travel time. Path is found using Dijkstra's algorithm based on both heuristic weights of edges and estimated travel times of edges as weights. Results show that paths computed using travel time as weights of edges have lesser total path cost than that of obtained by heuristic weights. 

In this work, the cost of traversing every edge is estimated, which facilitates to apply deterministic path planning algorithms like Dijkstra's algorithm, Bellmont-Ford algorithm \textit{et~cetera}. Many industries (like BlueBotics \cite{Blue:2009}) use topology maps to describe the floor and employs a depth-first search to generates a length-optimal path using deterministic path planning algorithms. This work is complementary to this approach where the travel times can be used as path determining factor in those deterministic algorithms without changing any model of computation or architecture.

The approach used in single-task case in this work can be extended in multi-task scenarios for a MR, where cost coefficient for different tasks has to be found out. This is a direction for future consideration and it could be extended to every MR in the system. During the run-time of MRS, every estimated value of travel time has context depending on various environmental and inherent factors. Travel time of one MR can provide contextual information to other MRs in an multi-robot system (MRS) and contribute in estimating travel time for them. This enhances further investigation towards implementing collaborative or collective intelligence in MRS to have cost efficient coordination of the MRS. 
\ifCLASSOPTIONcaptionsoff
  \newpage
\fi
\bibliographystyle{plain}
\bibliography{DR_jrnl_IEEE.bib}
\end{document}